\documentclass[]{article}

\usepackage{natbib}
\usepackage{graphicx}
\usepackage{dcolumn}
\usepackage{bm}
\usepackage{amssymb}
\usepackage{amsmath}
\usepackage{setspace}

\begin{document}

\begin{titlepage}
\begin{centering}
\begin{huge}
Coherently enhanced radiation reaction effects in laser-vacuum acceleration of electron bunches\\
\end{huge}
\begin{large}
\vspace{15mm}
P.W. Smorenburg$^1$, L.P.J. Kamp$^1$, G.A. Geloni$^2$, and O.J. Luiten$^1$\\
\end{large}
\begin{itshape}
\vspace{2mm}
$^1$Eindhoven University of Technology, Department of Applied Physics,\\
P.O. Box 513, 5600 MB, Eindhoven, The Netherlands\\
$^2$European XFEL GmbH,\\
Albert-Einstein-Ring 19, 22761 Hamburg, Germany\\
\end{itshape}
\vspace{40mm}
\end{centering}
Corresponding author:\\
P.W. Smorenburg,\\Eindhoven University of Technology, Department of Applied Physics, CQT Group\\
P.O. Box 513, 5600 MB, Eindhoven, The Netherlands\\p.w.smorenburg@tue.nl, tel. +31 40 2474030 or +31 40 2474359, fax +31 40 2438060 \\
\vspace{5mm}\\
Short title: "Coherently enhanced radiation reaction effects"\\
\vspace{5mm}\\
Number of manuscript pages: 31\\
Number of tables: 0\\
Number of figures: 4\\
\end{titlepage}

\begin{centering}
\begin{huge}
Coherently enhanced radiation reaction effects in laser-vacuum acceleration of electron bunches\\
\vspace{10mm}
\end{huge}
\end{centering}

\begin{abstract}
\doublespacing
The effects of coherently enhanced radiation reaction on the motion of subwavelength electron bunches in interaction with intense laser pulses are analyzed. The radiation reaction force behaves as a radiation pressure in the laser beam direction, combined with a viscous force in the perpendicular direction. Due to Coulomb expansion of the electron bunch, coherent radiation reaction effects only occur in the initial stage of the laser-bunch interaction while the bunch is still smaller than the wavelength. It is shown that this initial stage can have observable effects on the trajectory of the bunch. By scaling the system to larger bunch charges, these effects may be increased to such an extent that they can suppress the radial instability normally found in ponderomotive acceleration schemes, thereby enabling the full potential of laser-vacuum electron bunch acceleration to GeV energies.
\end{abstract}

\vspace{10mm}


\clearpage

\section{Introduction} \thispagestyle{empty} \doublespacing
The recent availability of ultra-intense laser pulses has led to increased efforts to develop novel compact acceleration schemes for electron beams. These include wakefield accelerators \citep{Tajima:1,Leemans:1,Malka:1,Jaroszynski:1}, thin foil irradiation schemes \citep{Maksimchuk:1,Borghesi:1} and laser-vacuum acceleration concepts \citep{Zolotorev:1,Wang:1}. Most studies so far have concentrated on the dynamics of the individual electrons in the considered electron bunches under the action of the externally applied electromagnetic fields, while consideration of the self-interaction of the bunches is limited to space-charge effects. Self-interaction due to the radiation emitted by the bunch itself is often ignored. Neglect of this collective radiation reaction is well-justified if the electron density is not too high and the electron bunch size exceeds the laser wavelength used.\\
In the last years, however, the interaction of intense laser light with nanometer to micrometer sized atomic clusters of near solid state density has become the subject of thorough investigation \citep{Saalmann:1,Krainov:1}. This has led to the observation of large numbers of electrons emitted from such clusters \citep{Shao:1,Springate:1,Fennel:1,Chen:1}. In particular, under suitable conditions the production of dense, attosecond electron bunches from laser-irradiated clusters has been observed both numerically \citep{Bauer:1} and experimentally \citep{Fennel:1,Fukuda:1}. In view of these developments, collective radiation reaction effects may become an important factor in the dynamics of dense electron bunches and should be taken into account. It was realized already many years ago \citep{Veksler:1} that these effects can even dominate the bunch dynamics and may be exploited as a collective acceleration mechanism. More recently, the ultra-intense field regime in which radiation reaction effects become dominant has been analyzed \citep{Bulanov:1} and the acceleration of plasma sheets assisted by radiation reaction has been considered \citep{Ilin:1,Kulagin:1} in the context of current technological possibilities.\\
In the present paper, we consider the effect of collective radiation reaction on a dense sub-wavelength electron bunch in interaction with a laser pulse. It is shown that the interaction can be modeled by a particularly simple picture of radiation pressure in the direction of the laser beam on a coherently enhanced effective cross section, in combination with a viscous force perpendicular to the laser propagation direction. Thus collective radiation reaction may be exploited in acceleration schemes as an additional accelerating force as well as a stabilizing force in the transverse direction.\\
\\
Classically, the net radiative effect of the interaction of charged particles with an electromagnetic wave is described by Thomson scattering. This is the production of secondary dipole radiation by charges that oscillate due to the electric field of an incident radiation wave, which can be seen as scattering of part of the incident radiation by the charges. In case of an electron bunch of charge $q=Ne$ and radius $R=R_0$ much smaller than the wavelength $\lambda$ of the incident radiation, the $N$ electrons in the bunch will essentially perform an identical oscillation, yielding coherently amplified secondary radiation that is identical to the radiation a point charge of magnitude $q$ would produce. Hence the dynamics of the bunch as a whole can be modeled by such a point charge, as long as $R\ll \lambda$ is fulfilled. The classical Thomson cross section, defined as the ratio of scattered power to incident intensity $I$ in the non-relativistic limit, is  $\sigma_T= N^2(8\pi/3)r_0^2$, with $r_0=e^2/(4\pi\epsilon_0m_ec^2)$ the classical electron radius \citep{Landau:1}. The factor $N^2$ shows the coherent enhancement of the cross section compared to that of a single electron. Since directed momentum of the incident radiation is scattered into dipole radiation with zero net momentum, momentum is transferred to the electron bunch at a rate $\sigma_T I/c\equiv F_T$. This means that the effective force per electron $F_T/N$ is proportional to $N$, and can therefore become significant for dense bunches. It is instructive to compare $F_T$ to the ponderomotive force $\bm{F}_P=-Ne^2\nabla I/(2\epsilon_0m_ec\omega^2)$ in a laser pulse, which is conventionally used to accelerate charges in laser-vacuum acceleration schemes. Using a typical laser pulse with $\lambda=1$ $\mu$m and 100 fs pulse length, the ratio $F_T/F_P$ becomes of the order of unity for a bunch of $N=10^6$ electrons. Assuming a laser intensity of $I=1\cdot10^{19}$ $\text{W}/\text{cm}^2$, this corresponds to a force of $F_T=22$ mN, which is equivalent to an electric field of $F_P/(Ne)=0.14$ TV/m in the laser propagation direction. This field already compares to the accelerating fields produced in wakefield accelerators, and increases to well above 1 TV/m for larger numbers of electrons.\\
Hence, radiation reaction is important for bunches containing $\gtrsim10^6$ electrons within a size $\ll\lambda$, that is, for bunches with a charge density close to solid state density. These electron bunches could only recently be extracted from atomic clusters by irradiation with an intense laser pulse. Of course, such dense electron bunches quickly expand beyond $R=\lambda$ due to the strong Coulomb repulsion, so that they will scatter radiation coherently only for a brief period of time after creation. Therefore coherent Thomson scattering will not in itself be an efficient driving mechanism for electron bunch acceleration. However, it will be shown in this paper that even a short initial period of radiation reaction dominated interaction can have an observable effect. For larger electron bunches containing more charge, the radiation reaction effects are correspondingly stronger, and may be increased to such an extent that they can suppress the radial instability normally found in ponderomotive acceleration schemes, thereby enabling the full potential of laser-vacuum electron bunch acceleration to GeV energies.\\

Collective radiation reaction of subwavelength electron bunches is, in some respects, the coherently enhanced version of the radiation reaction of a point charge. The latter has been a subject fundamental to the understanding of elementary particles and early electron theory and played a role in the development of quantum theory \citep{Fritz:1,Lorentz:1,Abraham:1,Dirac:1}. It has always been a purely academic subject due to the smallness of the effect, however, and is only now coming within experimental reach due to the availability of ultra-intense fields \citep{Bulanov:1}. We suggest that an interesting alternative may now be offered by electron bunches of subwavelength size. As explained above, for bunches the enhanced radiation reaction effects are much stronger so that they are accessible already using moderate laser intensities. Furthermore, also the characteristic time scale on which radiation reaction effects play a role (as measured by the quantity $\tau_e$ introduced below) is coherently prolonged, bringing these effects into the experimental attosecond regime. Thus the coherent enhancement of radiation reaction by electron bunches yields interesting possibilities to investigate this fundamental subject by experimental techniques.\\

In sections \ref{sec2}-\ref{sec4}, the bunch dynamics are analyzed treating the radiation reaction force in a perturbative way. In section \ref{sec2}, multiple-scale analysis and averaging over the laser optical time scale are applied to the equation of motion including radiation reaction, yielding a cycle-averaged description of the bunch dynamics. The results are applied to the case of a laser pulse in section \ref{sec3}, validating the averaged description and our numerical calculations. Subsequently, in section \ref{sec4} the theoretical results are placed in an experimental context by exploring the possibility of having subwavelength electron bunches to begin with, addressing both the time in which electron bunches Coulomb expand beyond wavelength and the production of dense bunches in laser-cluster interactions. Using accordingly realistic electron bunches, the observable effects of coherent radiation reaction on the bunch trajectory are calculated. Finally, in section \ref{sec5} the theoretical and numerical results are tentatively extrapolated to the non-perturbative regime of highly charged electron bunches. It is shown that ponderomotive acceleration of electron bunches by laser pulses may be stabilized by radiation reaction.\\

\section{Averaged radiation reaction force \label{sec2}}
Suppose that an electron bunch of charge $q=Ne$, mass $m=Nm_e$ and radius $R_0\ll \lambda$ oscillates due to the Lorentz force $\bm{F}_L$ of plane wave radiation propagating in the $z$-direction with electric field $\bm{E}=E_0\sin\left(\omega t-kz\right)\bm{e}_x$. The oscillatory quiver motion of the bunch (henceforth modeled by a point charge $q$) is a 'figure-of-eight'-cycle in the $\left(x,z\right)$-plane in the Lorentz frame in which the charge is at rest on average \citep{Landau:1}. In the course of each cycle, the accelerating charge continuously radiates and exchanges energy with its periodically changing Coulomb field, thereby experiencing an additional radiation reaction force $\bm{F}_R$. The detailed description of $\textbf{\emph{F}}_R$ and the associated equation of motion for a point charge have caused a long-standing debate \citep{Rohrlich:1}, in which the covariant equation of motion \citep{Lorentz:1,Abraham:1,Dirac:1},

\begin{equation}
\dot{p}^\mu=F_L^\mu+F_R^\mu=qF^{\mu\nu}\frac{p_\nu}{m}+\tau_e\left(\dot{p}^\lambda\dot{p}_\lambda p^\mu/(mc)^2+\ddot{p}^\mu\right) \label{1},
\end{equation}

has played a central role. In eq. (\ref{1}), dots denote differentiation with respect to proper time, $F^{\mu\nu}$ is the electromagnetic field tensor of the incident radiation, $\tau_e=N(2r_0)/(3c)$ is the characteristic time for radiation reaction, and the metric signature $(+,-,-,-)$ has been adopted. The first term in brackets equals the covariant Larmor power and represents the recoil of the charge caused by emission of radiation, that is, by the irreversible loss of four-momentum detaching from the charge and propagating towards infinity. The second terms is referred to as the 'Schott term' and is associated with the reversible exchange of four-momentum with the Coulomb or velocity field, that is, with the part of the electromagnetic field bound to the charge that does not give rise to radiation \citep{Teitelboim:1,Heras:1}. In the usual incoherent case $N=1$, the bracketed terms can be neglected because of the smallness of $\tau_e$. However, in the present case the radiation reaction parameter is $N$ times larger and $p^\mu\propto m\equiv Nm_e$, so that $F^\mu_R\propto N^2$. This shows again that the radiation reaction force is coherently enhanced and can become very relevant. Two regimes may be identified: the perturbative regime in which $\omega\tau_e\ll1$ and the radiation reaction dominated regime in which $\omega\tau_e\gtrsim1$ (see also \citep{Bulanov:1}). In the perturbative regime, the Thomson cross section is by definition much smaller than the physical bunch size, and $F_R\ll F_L$. This is the regime studied in this and the next two sections. In section \ref{sec5}, laser-vacuum acceleration in the radiation reaction dominated regime is considered. In passing, we note that eq. (\ref{1}) is a fundamental equation of motion for a structureless point charge, so that one must be careful using it to describe a finite charge distribution, be it much smaller than the wavelength of the problem. In particular, problems of interpretation occur when the mass equivalent of the electrostatic energy contained in the bunch becomes comparable to the rest mass $m=Nm_e$ \citep{Yaghjian:1}. This is not the case as long as $\omega\tau_e\ll kR_0$, however, which is always satisfied in the perturbative regime.\\

The two-dimensional 'figure-of-eight'-motion in the $(x,z)$-plane due to the linearly polarized incident radiation does not induce any $y$-component of acceleration in eq. (\ref{1}), so that the motion of the charge remains two-dimensional throughout. Assuming $F_R\ll F_L$, the radiation reaction force $\bm{F}_R$ affects the quiver motion negligibly on time scales $\sim \omega^{-1}$. However, its cumulative effect over many cycles is to accelerate the charge, so that it gains a slowly varying average momentum $\overline{\bm{p}}$ superimposed on the quiver motion. To study $\overline{\bm{p}}$, it is instructive to derive averaged equations of motion from eq. (\ref{1}) by means of a multiple-scale expansion. For this purpose, define a fast dimensionless time scale $T\equiv\omega t$ and a slow dimensionless time scale $\epsilon T\equiv (\omega\tau_e)T$. The latter equals the time scale of radiative damping of a charged harmonic oscillator \citep{Panofsky:1} so that this is the time scale on which $\overline{\bm{p}}$ will change appreciably. Hence, the dimensionless momentum $(\gamma,P_x,P_y,P_z)=P^\mu\equiv p^\mu/(mc)$ may be written as the sum of a slowly varying part $\overline{P}^\mu(\epsilon T)\equiv\langle P^\mu\rangle$ and a rapidly varying part $\tilde{P}^\mu(\epsilon T,T)=P^{(0)\mu}(\epsilon T,T)+\epsilon P^{(1)\mu}(\epsilon T,T)+O(\epsilon^2)$, where $\langle\cdot\rangle$ denotes the average over an optical period. Incidentally, the quiver motion in a plane wave turns out to be more easily described in terms of the laser phase $\eta\equiv T-kz$ rather than the time $T$ explicitly \citep{Landau:1}. Anticipating this fact by changing the fast coordinate from $T$ to $\eta$, time-differentiation takes the form

\begin{equation}
\gamma \frac{d}{dT}P^\mu(\epsilon T,\eta)=\left(\epsilon\gamma\frac{\partial}{\partial(\epsilon T)}+(\gamma-P_z)\frac{\partial}{\partial\eta}\right)P^\mu(\epsilon T,\eta), \label{1a}
\end{equation}

where $\gamma^2\equiv 1+P_x^2+P_y^2+P_z^2$ is the usual Lorentz factor. Substituting the sum $P^\mu=\overline{P}^\mu+\tilde{P}^\mu$ in the space and time components of eq. (\ref{1}), using (\ref{1a}), and collecting equal powers of $\epsilon$, yields the zeroth-order equations

\begin{eqnarray}
P_x^{(0)}&=&-a\cos\eta; \label{2}\\
P_z^{(0)}=\gamma^{(0)}&=&-\frac{a\overline{P}_x\cos\eta}{\overline{\gamma}-\overline{P}_z}+\frac{a^2\cos2\eta}{4\left(\overline{\gamma}-\overline{P}_z\right)};\label{3} \\
\overline{\gamma}^2&=&1+a^2/2+\overline{P}_x^2+\overline{P}_z^2, \label{4}
\end{eqnarray}

where $a\equiv(eE_0)/(m_ec\omega)$ is the dimensionless field strength. Eqs. ({\ref{2}-\ref{3}) give the well-known plane wave quiver motion with average momentum $\overline{\bm{p}}$ \citep{Salamin:2}. Eq. (\ref{4}) shows that the oscillating charge behaves like a particle with effective mass $m\sqrt{1+a^2/2}$ due to the energy contained in the quiver motion \citep{Kibble:1}. Expanding next the components of eq. (\ref{1}) to first order in $\epsilon$, substituting the zeroth-order results Eqs. (\ref{2}-\ref{4}), and averaging over the fast time scale, one finally obtains

\begin{eqnarray}
\frac{d\overline{P}_x}{dT}&=&-\frac{\omega\tau_e a^2}{2}(\overline{\gamma}-\overline{P}_z)^2\frac{\overline{P}_x}{\overline{\gamma}};\label{5}\\
\frac{d\overline{P}_z}{dT}&=&\frac{\omega\tau_e a^2}{2}(\overline{\gamma}-\overline{P}_z)^2\left(1-\frac{\overline{P}_x^2}{\overline{\gamma}\left(\overline{\gamma}-\overline{P}_z\right)}\right).\label{6}
\end{eqnarray}

The averaged dimensionless position $\bm{\overline{X}}\equiv k\bm{\overline{x}}$ is calculated from

\begin{equation}
\frac{d(\overline{\bm{X}})}{dT}\equiv\overline{\bm{\beta}}=\frac{\overline{\bm{P}}}{\overline{\gamma}}, \label{7}
\end{equation}

Eqs. (\ref{4}-\ref{7}) now give the evolution of the averaged position $\overline{\bm{X}}$ and momentum $\overline{\bm{P}}$ of the charge in the plane wave. First, consider the special case $\overline{P}_x=0$. Then Eqs. (\ref{5}-\ref{6}) reduce, after multiplication by $m_ec\omega$, to

\begin{equation}
\frac{d\bm{\overline{p}}}{dt}=\sigma_T\frac{I}{c}\overline{\gamma}^2\left(1-\overline{\beta}_z\right)^2\bm{e}_z, \label{7a}
\end{equation}

where $I=\epsilon_0cE_0^2/2$ is the intensity of the incident plane wave. This clearly shows that the averaged reaction force takes the classical form of a radiation pressure $I/c$ on an effective cross section $\sigma_T$, corrected by a Doppler factor $\overline{\gamma}^2\left(1-\overline{\beta}_z\right)^2$. This Doppler factor reflects the fact that plane wave appears Doppler-shifted in the frame of the charge, with a corresponding change in radiation pressure. Eq. (\ref{7a}) may also be obtained by Lorentz-transformation of the non-relativistic force $\sigma_T I/c$ \citep{Landau:1}. In the more general case $\overline{P}_x\neq0$, the direction of the radiation pressure is no longer along the propagation direction of the incident wave. This is because the secondary, scattered radiation produced by the charge is Lorentz boosted along $\overline{\bm{P}}$, so that the recoil from this radiation has a net transverse component when $\overline{P}_x\neq 0$. The latter appears in Eqs. (\ref{5}-\ref{6}) as the additional factors involving $\overline{P}_x$ on the very right. Effectively, in the transverse direction the averaged reaction force appears as a frictional force proportional to the velocity, which will be important below. Effective forces similar to Eqs. (\ref{5}-\ref{6}) have been obtained earlier for a one-dimensional plasma sheet \citep{Ilin:1}. \\

A typical experiment in which the time-averaged forces exhibited by Eqs. (\ref{5}-\ref{6}) are important will involve the use of a high-intensity laser pulse of length $\tau_L$ focused to a waist of size $w_0$. The laser pulse will appear to the electron bunch as a slowly modulated plane wave if $c\tau_L\gg\lambda$ and $w_0\gg k^{-1}$. The averaged reaction force will be as in Eqs. (\ref{5}-\ref{6}) with the dimensionless field strength $a(\bm{\overline{X}},T)$ now evolving according to the laser pulse shape. In this setting, however, the electron bunch will in addition experience a ponderomotive force due the the presence of large intensity gradients in the laser pulse. This is an additional average force which should be properly added to the right hand sides of Eqs. (\ref{5}-\ref{6}). It can be shown, by a multiple-scale analysis analogous to the one described above, that the ponderomotive force can be accurately described by

\begin{equation}
\frac{\bm{F}_P}{m_ec\omega}=-\frac{(k^{-1}\nabla) [a^2(\overline{\bm{X}})]}{4\overline{\gamma}} \label{9}
\end{equation}

if $\lambda/(c\tau_L)\sim (kw_0)^{-1}\ll1$ \citep{Quesnel:1}. When adding $\bm{F}_P$ to the right hand sides of Eqs. (\ref{5}-\ref{6}), in effect two separate multiple-scale analyses are combined. This is only correct provided that the small parameters used in both analyses are of the same order, which in this case means that also $\omega\tau_e\sim \lambda/(c\tau_L)\sim (kw_0)^{-1}$.\\

Rather than using averaged equations, the detailed trajectory of the electron bunch may be calculated by direct integration of Eq. (\ref{1}). However, it is well-known that this equation yields unphysical runaway solutions \citep{Jackson:1}, which may be disposed of by substituting the applied Lorentz force $\dot{p}^\mu\approx F^\mu_L$ in the right hand side of Eq. (\ref{1}). This yields the space component \citep{Landau:1}

\begin{equation}
\frac{d\bm{p}}{dt}=\bm{F}_L+\tau_e\left[\gamma\frac{d\bm{F}_L}{dt}-\frac{\gamma^3}{c^2}\left(\frac{d\bm{v}}{dt}\times\left(\bm{v}\times\bm{F}_L\right)\right)\right]. \label{8}
\end{equation}

This expression can also be derived from quantum mechanical arguments \citep{Ford:1}, and has been put forward as the exact equation of motion rather than Eq. (\ref{1}) \citep{Rohrlich:2}.

\section{Radiation reaction in a laser pulse \label{sec3}}
In this section the results of the previous section will be applied to concrete examples of laser-vacuum experiments. First, however, it should be stressed that an accurate calculation of the detailed charge trajectory in a laser pulse requires the field of the pulse to be described in more detail than the usual paraxial approximations \citep{Hora:1}, even in the absence of radiation reaction forces. In \citep{Quesnel:1}, a laser field representation in Fresnel-type integrals is used to calculate detailed charge trajectories in a laser pulse, which are then compared with averaged trajectories calculated from the ponderomotive force Eq. (\ref{9}) using the laser field in paraxial approximation. The agreement between the two results was found excellent except for highly relativistic initial velocities. In this paper we will adopt the same approach, using the detailed field representation of \citep{Quesnel:1} when evaluating the Lorentz force $\bm{F}_L$ in Eq. (\ref{8}), while taking the simpler paraxial approximation to evaluate the field intensity $a^2$ in the averaged equations (\ref{4}-\ref{9}). As a reference, both field representations are shown in the appendix.\\

To validate the results of the previous section and to illustrate the effect of the effective radiation pressure produced by radiation reaction, we consider an intense laser pulse propagating in the $z$-direction and polarized in the $x$-direction, which has wavelength $\lambda= 1$ $\mu$m, peak intensity $I=1\cdot10^{19}$ W/cm$^2$ and pulse length $\tau_L=200$ fs, focused to a relatively large waist of size $w_0=10\lambda$. This pulse contains $2.3$ J of energy and has a peak dimensionless field strength $a_{max}=2.7$. An electron bunch of radius $R_0\ll \lambda$ and radiation reaction parameter $\omega\tau_e=0.03$ is considered, corresponding to a bunch charge of $q=0.4$ pC. For these parameters, $\lambda/(c\tau_L)\sim (kw_0)^{-1}\sim\omega\tau_e$ so that the averaged description should be valid. The electron bunch is placed at rest at the initial position $\bm{x}_0\equiv(x_0,z_0)=(\lambda,0)$ prior to arrival of the laser pulse, deliberately at an off-axis position arbitrarily set to one wavelength, in order to study the radial acceleration as well. In this (unrealistic) first example, we disregard the Coulomb expansion of the bunch and assume that it retains its coherent Thomson cross section during the entire interaction, in order to study the basic equations governing the radiation reaction effects.\\

Fig. \ref{fig1} shows the trajectory of the bunch and its momentum as a function of time as it is overtaken by the laser pulse, calculated using both the averaged description Eqs. (\ref{4}-\ref{9}) and the detailed description Eq. (\ref{8}). Clearly, the agreement between the averaged and the detailed description is excellent, validating both the multiple-scale analysis and our numerical calculation. The figure also shows the trajectory and the momentum in the absence of radiation reaction, that is, in case of conventional ponderomotive acceleration. In the latter case the charge is quickly expelled in the radial direction because of the large radial intensity gradient, which is a well-known instability that so far has limited the applicability of ponderomotive acceleration schemes \citep{Zolotorev:1}. The theoretical escape angle $\theta$ for ponderomotive acceleration by a radiation pulse of a charge initially at rest, is related to the final Lorentz factor $\gamma_f$ by \citep{Hartemann:1}

\begin{equation}
\theta=\arccos\sqrt{\frac{\gamma_f-1}{\gamma_f+1}}. \label{8a}
\end{equation}

In the present example $\gamma_f=1.96$, yielding a theoretical escape angle of $55.3^o$ in excellent agreement with the numerically calculated result. In contrast to the purely ponderomotive case, in the case including coherent Thomson scattering the acceleration in the radial direction is markedly suppressed. This stabilizing effect is due to the average frictional force shown by Eq. (\ref{5}), which opposes the tendency to radial acceleration. Meanwhile, the charge stays in the beam longer and is accelerated in the forward direction for a longer time by both the radiation reaction pressure and ponderomotive forces, leading to a higher final energy and a lower escape angle.\\

\section{Realistic electron bunches \label{sec4}}
We now consider the more realistic case of an electron bunch that expands due to Coulomb repulsion. The time after which the Thomson cross section fails to be coherent, that is, the time it takes the bunch to expand to $R=\lambda$, depends on the charge and the initial size of the bunch. To obtain realistic values for these initial parameters, we first make some estimates concerning the method to produce dense electron bunches mentioned in the introduction.\\
Rather than irradiating a pre-existing electron bunch, an electron bunch may be created in practice by irradiating a sub-wavelength atomic cluster. When the laser pulse is sufficiently strong, the electron bunch will be emitted from the cluster somewhere in the leading edge of the pulse, after which it is available to be accelerated by the remainder of the pulse as has been analyzed in the previous section. Many different mechanisms play a role in laser-cluster interaction \citep{Saalmann:1,Krainov:1} and it is beyond the scope of this paper to analyze this interaction in detail. Instead, we use a strongly simplified model of the cluster suggested by \citep{Parks:1} to obtain order-of-magnitude estimates for the initial parameters of the electron bunch. Accordingly, at some point in the leading edge of the pulse the field strength becomes such that all atoms in the cluster are ionized almost instantaneously, changing the neutral cluster into a dense plasma ball. Subsequently, this plasma ball is modeled as a spherical rigid electron bunch, interpenetrating with and moving through a practically immobile oppositely charged ion bunch, under the action of both the driving laser electric field and the restoring Coulomb force of the ion cloud. As the rising edge of the laser pulse advances, the electron bunch oscillates around the ion bunch center with increasing amplitude. At a critical laser field strength the electron bunch breaks free and escapes the ion bunch within a fraction of an optical period, yielding the free, dense electron bunch we will now use as an input for our calculation.\\
The critical dimensionless field strength at which the electron bunch is liberated may be estimated as \citep{Parks:1}

\begin{equation}
a_\text{crit}=\frac{15}{96}\left(\frac{\omega_p}{\omega}\right)^2kR_0, \label{10}
\end{equation}

where $\omega_p=\sqrt{3Ne^2/(4\pi\epsilon_0m_eR_0^3)}$ is the plasma frequency of the ionized cluster consisting of $N$ atoms. The electron bunch leaves the ion bunch at the maximum of an optical cycle. The time this takes is of the order $t_\text{esc}\sim\omega_p^{-1}$ \citep{Parks:1}, so that the escaping electron bunch will have a velocity of approximately

\begin{equation}
\beta_\text{esc}c\approx\omega_pR_0 \label{11}
\end{equation}

in the direction of laser polarization. Incidentally, note that the coherent radiation reaction parameter can be expressed in terms of the cluster parameters as $\omega\tau_e=(2/9)(\omega_p/\omega)^2(kR_0)^3$, by which the validity condition $\omega\tau_e\ll kR_0$ of section \ref{sec2} automatically implies that $\beta_\text{esc}< 1$. After leaving the cluster, the electron bunch will start to expand. It can be shown straightforwardly that, after a short period of slow expansion, the expansion rate increases to the constant value \citep{Ammosov:1}

\begin{equation}
\frac{dR}{dt}\approx\frac{\sqrt{\frac{2}{3}}\omega_pR_0}{\sqrt{1+P_\text{esc}^2+\frac{a_\text{crit}^2}{2}}}. \label{12}
\end{equation}

In Eq. (\ref{12}), it has been used that the expanding electron bunch will be in quiver motion immediately after leaving the cluster, so that it will move with an average Lorentz factor according to Eq. (\ref{4}). This reduces the expansion rate by a factor $\overline{\gamma}=\sqrt{1+P_\text{esc}^2+a_\text{crit}^2/2}$.\\

As an example we take an $R_0=16$ nm cluster with an electron density of $5\cdot10^{28}$ m$^{-3}$, which can be routinely made using a supersonic gas jet \citep{Krainov:1}, irradiated by the same laser pulse as in the previous example. With these numbers the other parameters are $kR_0=0.1$, $\omega_p/\omega=6.6$, $\omega\tau_e=0.01$, and $a_\text{crit}=0.68$ and $\beta_\text{esc}=0.66$ according to Eqs. (\ref{10}-\ref{11}). The last two numbers roughly compare to the field strengths used and electron velocities obtained in \citep{Bauer:1,Fukuda:1}. Assuming the expansion rate Eq. (\ref{12}), the bunch grows larger than the wavelength after a time $8.3\omega^{-1}$, that is, after no more than 1.3 optical cycles following the moment of escape from the cluster. Thus coherent Thomson scattering will only occur for a brief initial period. Nevertheless, even such a brief period can have an observable effect on the trajectory of the electron bunch in interaction with the laser pulse.\\
Fig. \ref{fig2} shows the trajectory of the bunch in interaction with the laser pulse, both for the case without the radiation reaction force ($\omega\tau_e=0$) and for the case including the radiation reaction force ($\omega\tau_e=0.01$) during the initial period $0\leq \omega t\leq 8.3$. In the latter case, the decrease in coherent radiation reaction was modeled by a sudden switch-off, setting $\omega\tau_e=0$ at the somewhat arbitrary point $\omega t=8.3$. Again, the results of Fig. \ref{fig2} were calculated both using the averaged description Eqs. (\ref{4}-\ref{9}) and the detailed description Eq. (\ref{8}). In both cases the initial position was $\bm{x}_0=(-\lambda,0)$ and the initial velocity was $\beta_\text{esc}c\bm{e}_x$. The initial position of the laser pulse along the $z$-axis was chosen such that $a=a_\text{crit}$ at $\bm{x}_0$, and the optical phase offset $\phi_0$ was chosen such that the optical cycle was at its maximum at $t=0$. This time, the escape angle for the case without radiation reaction does no longer agree with Eq. (\ref{8a}) or its generalization for a nonzero initial velocity \citep{Hartemann:1}. This may be caused by the sudden injection into the radiation field at $t=0$, which is not considered in the derivation of Eq. (\ref{8a}). The good agreement between the averaged and detailed descriptions is remarkable here, considering the fact that the concept of averaging the equation of motion over the optical time scale is not a priori valid when applied during the initial period, which is as short as the optical period itself.\\
From the figure it is clear that the initial period of radiation reaction still reduces the escape angle observably, be it much less than in the previous example. To put the $1.3$ degrees of reduction in angle in perspective, the figure also shows the error margins around the trajectory without radiation reaction which would be caused by an uncertainty of magnitude $\lambda$ in the initial position $\bm{x}_0$. The deviation caused by radiation reaction is clearly outside these error margins.\\

As another example we show how the initial period of coherent Thomson scattering may even change the behavior of the bunch completely. Fig. \ref{fig3} shows the trajectory of the bunch, using the same laser parameters and initial conditions as in the previous example, only now with the initial position changed to $\bm{x}_0=(-7.1\lambda,0)$. This time the effect of radiation reaction is such that the bunch is deflected by the laser pulse in the negative $x$-direction instead of the positive $x$-direction. Under the initial conditions of this example, the bunch is produced much further from the laser beam axis and is therefore decelerated in the negative $x$-direction by the radial ponderomotive force of Eq. (\ref{9}) for a longer time. Without the action of the radiation reaction force, the bunch just makes it across the laser axis (where the ponderomotive potential is maximum), after which it is accelerated in the positive $x$-direction. Including radiation reaction, however, in the initial stage the bunch is additionally decelerated by the frictional force of Eq. (\ref{5}). Furthermore it is additionally accelerated in the positive $z$-direction by the radiation pressure of Eq. (\ref{6}), so that it keeps up with the laser pulse and is under the action of the decelerating radial ponderomotive force for a somewhat longer time. Both effects are only small, but in this case just enough to prevent the bunch from passing the beam axis. It is to be noted that in this example whether or not the bunch will pass the beam axis is very sensitive to the initial parameters. This is an additional reason why the averaged description cannot be used here to reproduce the trajectories of Fig. (\ref{fig3}), since the small differences between the two descriptions in the initial part of the trajectory affect the final behavior strongly. Still, it is clear that radiation reaction can play an important role in laser-vacuum experiments.

\section{Pre-acceleration using radiation reaction \label{sec5}}
In the previous section, relatively small electron bunches were considered leading to modest values for the bunch charge and for the radiation reaction parameter ($\omega\tau_e\ll1$), thereby staying well inside the perturbative regime studied in sections \ref{sec2}-\ref{sec4}. Much stronger radiation reaction effects could be expected when scaling the system to larger bunches. In the introduction, the magnitude of the effective radiation reaction force was estimated assuming a wavelength of 1 $\mu$m. Taking now, for instance, a CO$_2$ laser with a wavelength of 10 $\mu$m, ten times larger bunches still Thomson scatter coherently. Since the coherent Thomson cross section scales as $\sigma_T\propto N^2\propto R^6$, in that case an effective force of $F_T=\sigma_T I/c\sim$ kN rather than mN may be expected, equivalent to accelerating fields approaching $F_P/(Ne)\sim$ PV/m rather than TV/m. Thus, the radiation reaction force in this case completely dominates the bunch dynamics. But as before, the action of the radiation reaction force is restricted to the initial period following the creation of the electron bunch, while the bunch is still smaller than the wavelength. Therefore the large accelerating fields just mentioned are not effective as a driving acceleration mechanism. However, the initial radiation reaction dominated phase pre-accelerates and redirects the velocity of the bunch, which may serve to stabilize subsequent ponderomotive acceleration.\\
As an example of such acceleration using radiation reaction, consider the same laser pulse as used in the previous examples, but scaled ten times in all directions so that $\lambda=10$ $\mu$m, $\tau_L=2$ ps and $w_0=10\lambda$. This CO$_2$ laser pulse contains $2.3$ kJ of energy and has a peak dimensionless field strength $a_{max}=27$, which is presently available. The increased wavelength allows an electron bunch of initial size $R_0=160$ nm containing $N=8.5\cdot10^8$ electrons, resulting as before in $kR_0=0.1$, but this time in a much bigger value $\omega\tau_e=1$. The non-relativistic cluster model of \citep{Parks:1} does not apply for such a large amount of charge. We assume a relativistic initial electron bunch velocity of $\beta_0=0.9$ instead, as is also indicated in \citep{Bauer:1}. Substituting $P_\text{esc}\equiv\beta_0/\sqrt{1-\beta_0^2}$ and $a_\text{crit}\equiv a_\text{max}$ in Eq. (\ref{12}), the resulting expansion rate is such that the bunch stays smaller than the wavelength for about one optical cycle, just as in the previous examples.\\
Fig. \ref{fig4} shows the trajectory and momentum of this electron bunch in the laser pulse just described. The initial bunch position was set to $\bm{x}_0=(\lambda,0)$, the initial position of the laser pulse along the $z$-axis was chosen centered around $x$-axis so that the bunch started its movement at the pulse maximum, and the optical phase offset $\phi_0$ was chosen such that the optical cycle was at its maximum at $t=0$ and $\bm{x}_0$. From the figure the effect of the initial period of radiation reaction is evident: the large radial velocity of the bunch is suppressed immediately preventing an early escape from the beam waist, after which the bunch is strongly accelerated in the direction of propagation of the laser pulse. The inset of the $p_z$-panel of Fig. 4 shows the long-term evolution of the momentum in this direction. The bunch is first accelerated by the leading edge of the laser pulse and then decelerated by the trailing edge, as is characteristic for laser-vacuum acceleration. The energy of the bunch is increased to $\gamma\approx600$, corresponding to an energy of $0.3$ GeV per electron.\\
In the case without radiation reaction, the bunch is quickly expelled from the laser beam. Consequently, the energy gained is much less. The cusp-like features in the corresponding $p_z$- and $p_x$-plots of Fig. \ref{fig4} agree with previous results \citep{Dodin:1,Galkin:1} in comparable cases of ponderomotive scattering.\\

Returning to the case including radiation reaction, it may seem strange that the bunch starts, at $t=0$, with a downward radial acceleration while the laser electric field points upward at that moment. This, however, is a consequence of the large value of $\omega\tau_e$ as can be inferred from Eq. (\ref{8}). Namely, at $t=0$ the optical field at the bunch position is maximum, so that

\begin{eqnarray}
q\bm{E}&=&\bm{e}_xm_ec\omega a_\text{max};\nonumber\\
qc\bm{B}&=&\bm{e}_ym_ec\omega a_\text{max};\nonumber\\
\frac{d\bm{F}_L}{dt}&=&q\frac{d\bm{v}}{dt}\times \bm{B},\nonumber
\end{eqnarray}

while $p_z=0$. Substituting the above in the $x$- and $z$-components of Eq. (\ref{8}), writing $d\bm{v}/dt$ in terms of the momentum $\bm{p}=\gamma m\bm{v}$, and rearranging, one finds in dimensionless form the $x$-component

\begin{equation}
\left.\frac{dP_x}{dT}\right|_{t=0}=a_\text{max}\frac{1-\omega\tau_ea_\text{max}\gamma_0^2\beta_0}{1+\gamma_0^2(\omega\tau_ea_\text{max})^2},\label{13}
\end{equation}

where $\gamma_0\equiv(1-\beta_0^2)^{-1/2}$. In the example of Figure \ref{fig4}, $\omega\tau_ea_\text{max}\beta_0\gg 1$, so that Eq. (\ref{13}) reduces to

\begin{equation}
\left.\frac{dP_x}{dT}\right|_{t=0}\approx-\frac{\beta_0}{\omega\tau_e}.\label{14}
\end{equation}

In the inset of the $p_x$-panel of Fig. \ref{fig4}, this initial acceleration is indicated by the sloped line. Eq. (\ref{13}) suggests that the initial velocity of the bunch can be redirected by means of a short period of radiation reaction choosing favorable parameters $a_\text{max}$ and $\omega\tau_e$, and may be used to efficiently accelerate the bunch into the direction of propagation of the laser such as has been done in the example of Fig. \ref{fig4}.\\
It should be noted that the very large radiation reaction parameter used in the last example poses some challenges to the theory of this paper. In particular, here $\omega\tau_e\gg kR_0$, which makes the applicability of the point charge equation of motion (\ref{1}) to an extended electron bunch debatable. Furthermore, Eq. (\ref{8}) has been proposed on the one hand as an approximation to the classical equation of motion (\ref{1}) based on the assumption $\omega\tau_e\ll1$ \citep{Landau:1}, and on the other hand as the exact equation of motion replacing Eq. (\ref{1}) for all values of $\omega\tau_e$ \citep{Rohrlich:2}. Interestingly, the experimental availability of charged objects for which $\omega\tau_e\sim1$ may now give some insights about which approach is right, and may offer further opportunities to experimentally test radiation reaction theories.

\section{Conclusion \label{sec6}}
The current developments in laser technology make it possible to obtain subwavelength electron bunches of very high charge density. With that, a qualitatively new regime is accessed in which coherently enhanced radiation reaction effects become significant. In this paper, we have analyzed some of these effects in the context of laser-vacuum experiments. It has been shown that the radiation reaction force affects the bunch dynamics notably, even if the radiation reaction can still be treated as a small perturbation on the optical time scale. Considering larger bunches containing more charge, we demonstrated that the radiation reaction effects may even become strong enough to be exploited in effective bunch acceleration schemes, although this is also where the theory needs to be further developed. It is clear that the coherently enhanced radiation reaction of high-density electron bunches offers interesting new possibilities, both as a technological tool in the development of novel acceleration schemes, and as an experimentally accessible system to study the fundamental topic of radiation reaction.

\section*{Appendix A: Laser field representations}
In this paper, when evaluating the Lorentz force $\bm{F}_L$ in Eq. (\ref{8}), the following fields are used \citep{Quesnel:1}:

\begin{eqnarray}
E_x&=&\frac{(kw_0)^2E_0}{4}\left(I_1+\frac{x^2}{kr^3}I_2\right);\label{A1}\\
E_z&=&\frac{(kw_0)^2E_0}{4}\frac{x}{r}I_4;\label{A2}\\
cB_y&=&\frac{(kw_0)^2E_0}{4}\left(I_1-\frac{x^2}{kr^3}I_2+\frac{x^2}{r^2}I_3\right);\label{A3}\\
E_y&=&cB_x=cB_z=0;\label{A4}\\
I_1&\equiv&\int_0^1 e^{-\frac{(kw_0)^2b^2}{4}}\left(1+\sqrt{1-b^2}\right)\sin\phi J_0(krb)bdb;\nonumber\\
I_2&\equiv&\int_0^1 e^{-\frac{(kw_0)^2b^2}{4}}\frac{1}{\sqrt{1-b^2}}\sin\phi J_1(krb)b^2db;\nonumber\\
I_3&\equiv&\int_0^1 e^{-\frac{(kw_0)^2b^2}{4}}\frac{1}{\sqrt{1-b^2}}\sin\phi J_0(krb)b^3db;\nonumber\\
I_4&\equiv&\int_0^1 e^{-\frac{(kw_0)^2b^2}{4}}\left(1+\frac{1}{\sqrt{1-b^2}}\right)\cos\phi J_1(krb)b^2db;\nonumber\\
\phi&\equiv&\omega t-kz\sqrt{1-b^2}+\phi_0,\nonumber
\end{eqnarray}

where $\phi_0$ is the phase offset of the optical field. These expressions describe a Gaussian beam which propagates along the $z$-direction, is polarized in the $x$-direction, and is focused to a waist of size $w_0$. To obtain a laser pulse of length $\tau_L$ instead of a full beam, the above fields are multiplied by an envelope function $\cos^2\left(\pi\eta/(2\omega\tau_L)\right)$. To evaluate the field intensity $a^2$ in the averaged equations (\ref{4}-\ref{9}), the simpler paraxial approximation is used according to

\begin{eqnarray}
a^2&=&\left[\frac{eE_0}{m_ec\omega}\frac{w_0}{w}e^{-\frac{w_0^2}{w^2}(kx)^2}\cos^2\left(\frac{\pi\eta}{2\omega\tau_L}\right)\right]^2;\label{A5}\\
w&\equiv&w_0\sqrt{1+4\frac{(kz)^2}{(kw_0)^4}}.\nonumber
\end{eqnarray}

\clearpage
\begin{figure}[p!]
\centering
\begin{tabular}{cc}
\includegraphics[width=10cm]{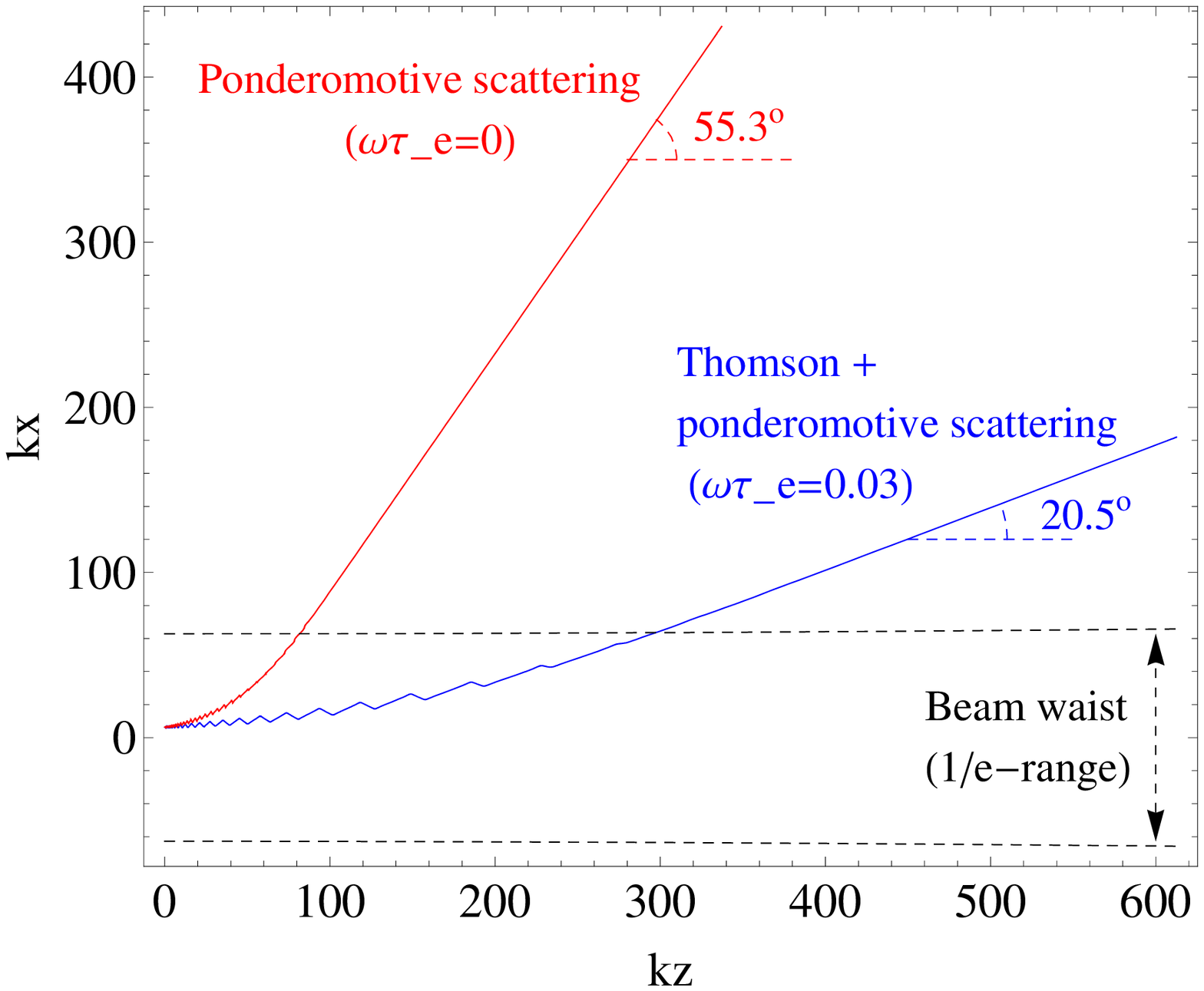}\\
\includegraphics[width=10cm]{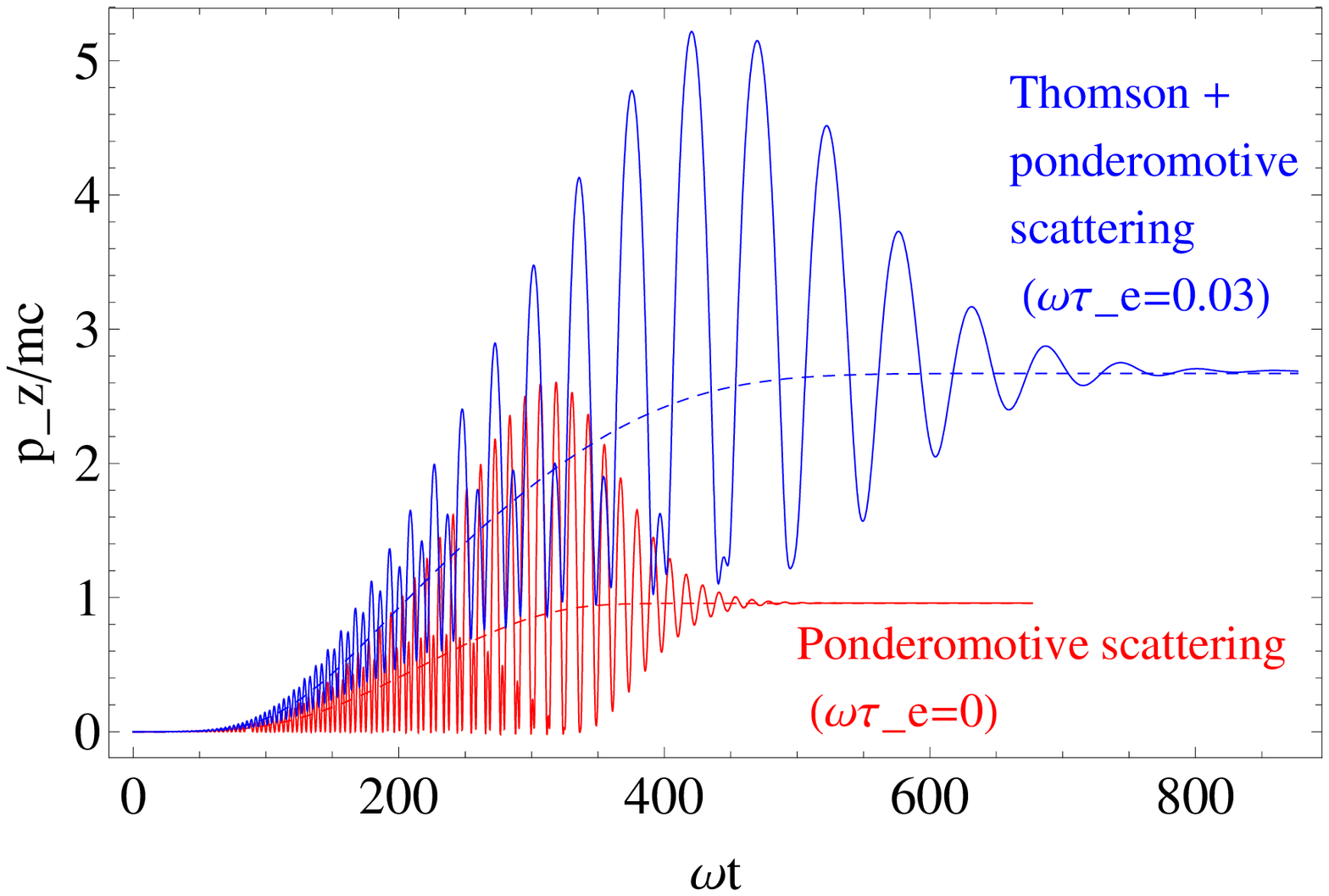}\\
\includegraphics[width=10cm]{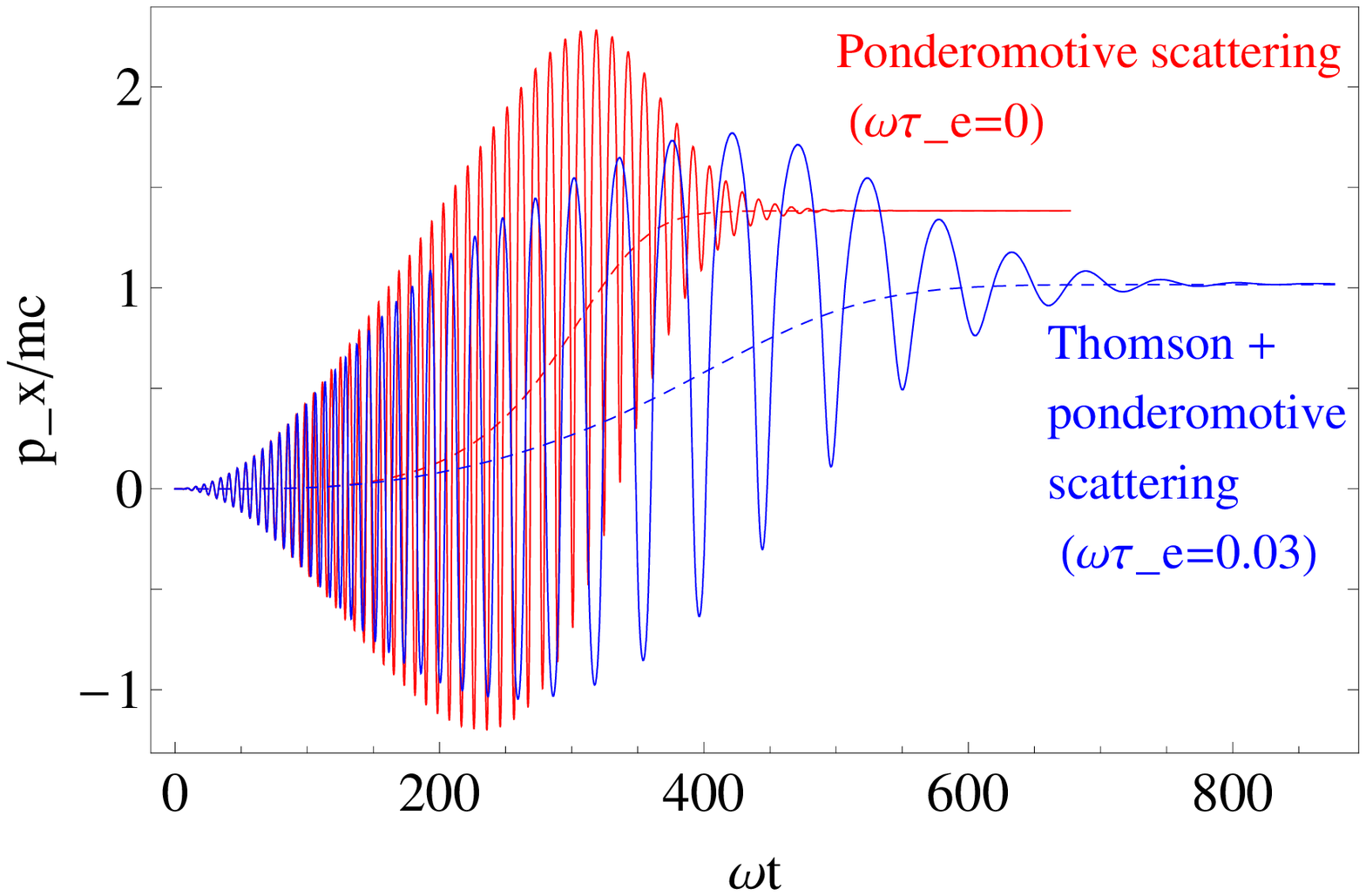}
\end{tabular}
\end{figure}

\clearpage
\begin{figure}[p!]
\caption{\label{fig1} \doublespacing
a) Trajectory of an electron bunch accelerated by a laser pulse with parameters $\lambda= 1$ $\mu$m, $I=1\cdot10^{19}$ W/cm$^2$, $\tau_L=200$ fs, and $w_0=10\lambda$; b) Longitudinal momentum of the bunch as a function of time; and
c) Transverse momentum of the bunch as a function of time, calculated for the case including radiation reaction (Thomson + ponderomotive scattering, $\omega\tau_e=0.03$) and for the case without radiation reaction (ponderomotive scattering, $\omega\tau_e=0$). The solid lines have been calculated using Eq. (\ref{8}) with $\bm{F}_L=q(\bm{E}+\bm{v}\times\bm{B})$, taking for $\bm{E}$ and $\bm{B}$ Eqs. (\ref{A1}-\ref{A4}). The dashed lines have been calculated using Eqs. (\ref{4}-\ref{9}), taking $a^2$ according to Eq. (\ref{A5}). The initial position was $\bm{x}_0=(\lambda,0)$ and the initial velocity was zero.
}
\end{figure}

\clearpage
\begin{figure}[p!]
\includegraphics[width=12cm]{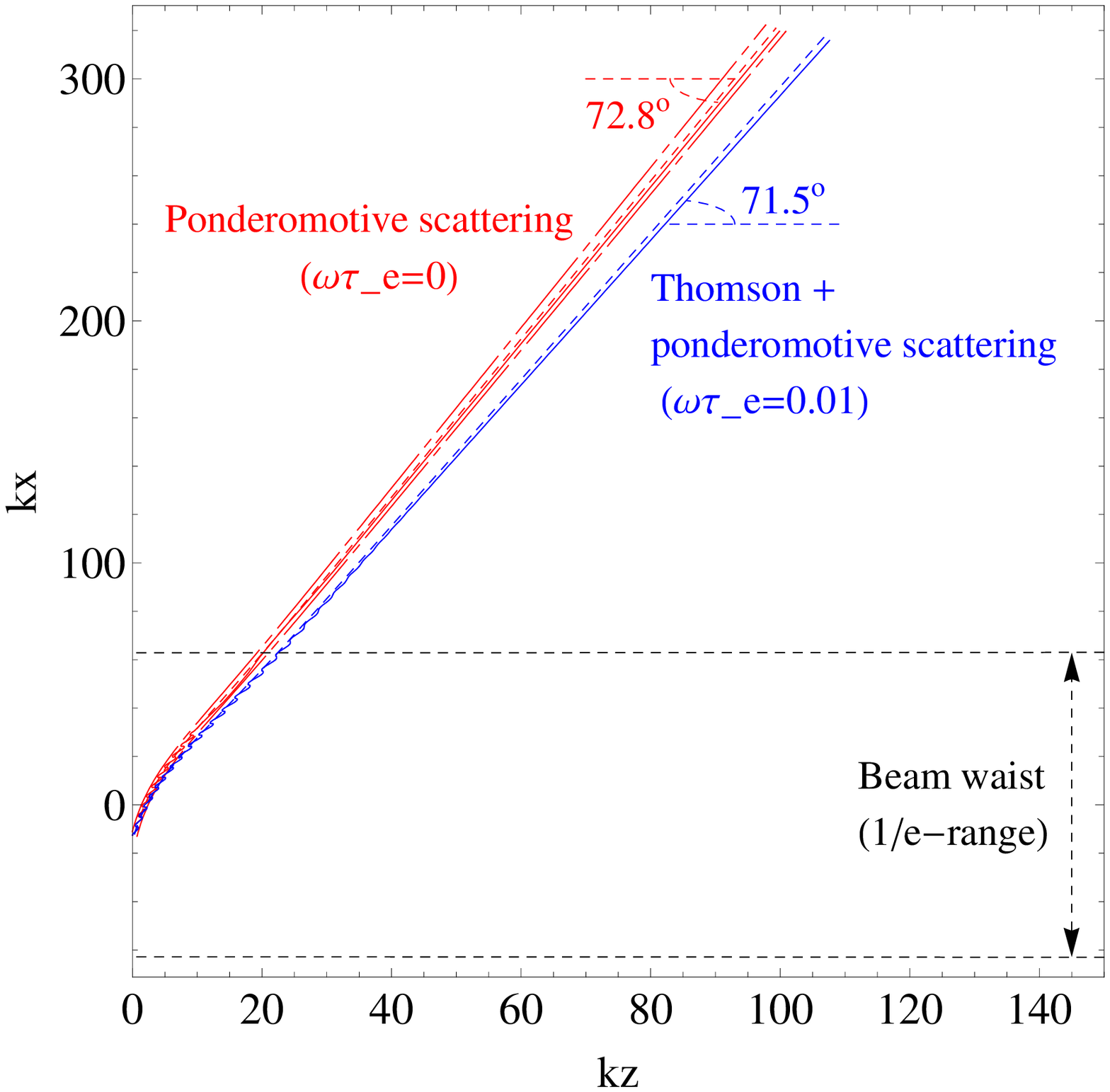}
\end{figure}

\clearpage
\begin{figure}[p!]
\caption{\label{fig2} \doublespacing
Trajectory of an electron bunch accelerated by a laser pulse, calculated for the case including radiation reaction (Thomson + ponderomotive scattering, $\omega\tau_e=0.01$) during the initial period $0\leq \omega t\leq 8.3$, and for the case without radiation reaction (ponderomotive scattering, $\omega\tau_e=0$). See Fig. (\ref{1}) for details about the calculation and laser parameters. The dash-dotted lines show the error margins for the case $\omega\tau_e=0$ corresponding to an hypothetical uncertainty of magnitude $\lambda$ in the initial position, calculated using the averaged description. The initial position was $\bm{x}_0=(-\lambda,0)$ and the initial velocity was according to Eq. (\ref{11}).
}
\end{figure}

\clearpage
\begin{figure}[p!]
\includegraphics[width=12cm]{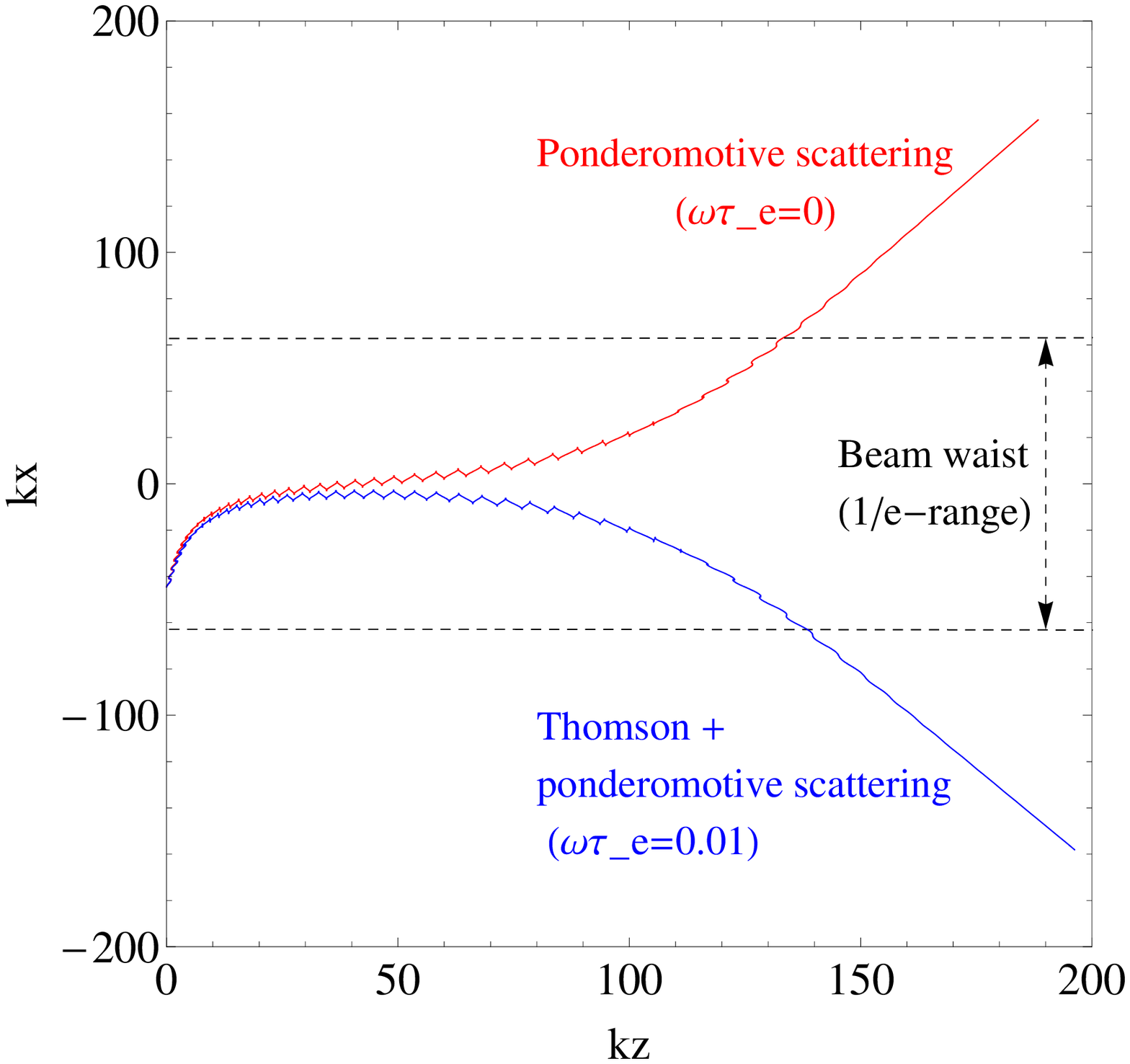}
\end{figure}

\clearpage
\begin{figure}[p!]
\caption{\label{fig3} \doublespacing
See Fig. (\ref{fig2}), with the initial position changed to $\bm{x}_0=(-7.1\lambda,0)$.
}
\end{figure}

\clearpage
\begin{figure}[p!]
\centering
\begin{tabular}{c}
\includegraphics[width=10cm]{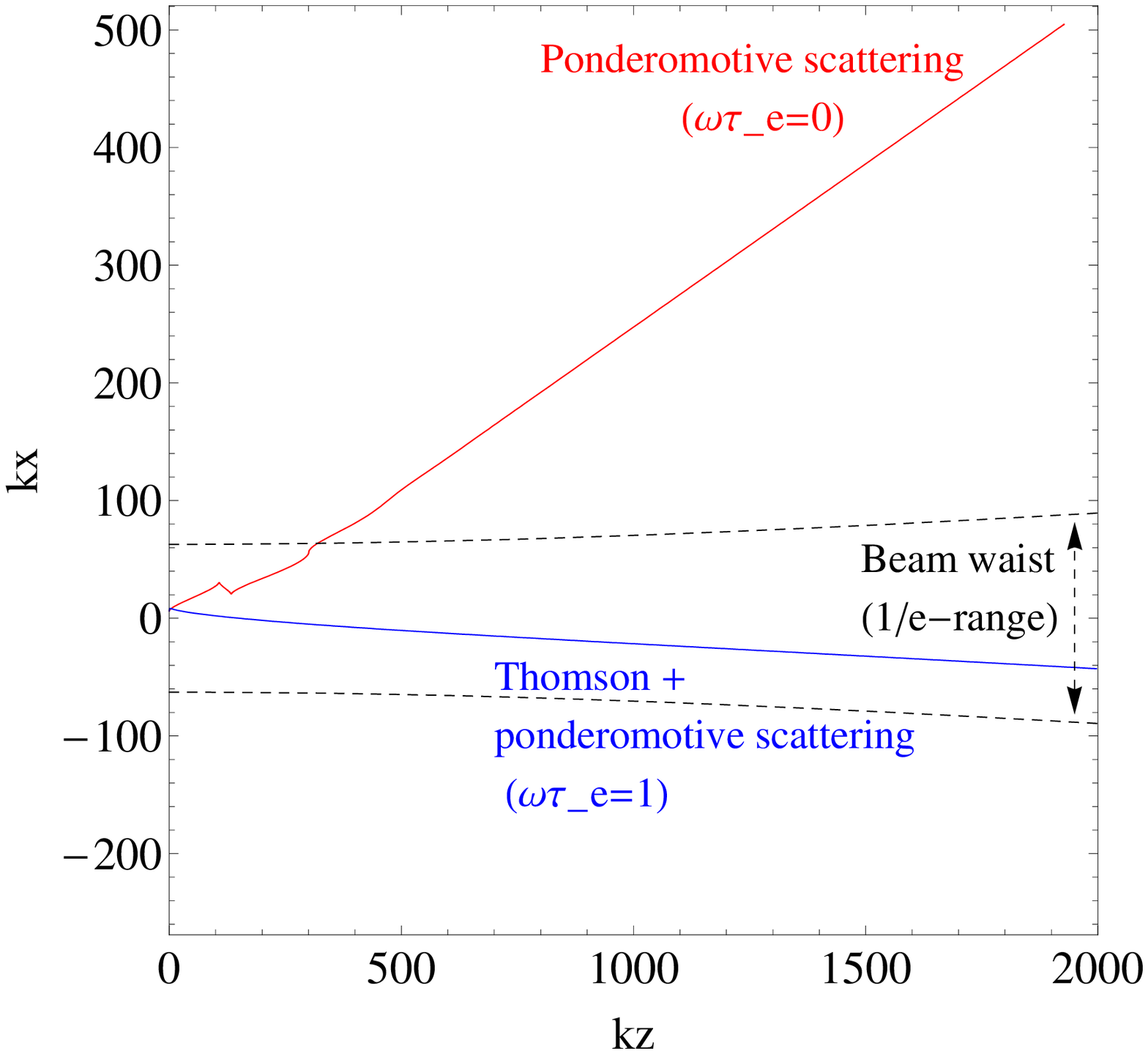}\\
\includegraphics[width=10cm]{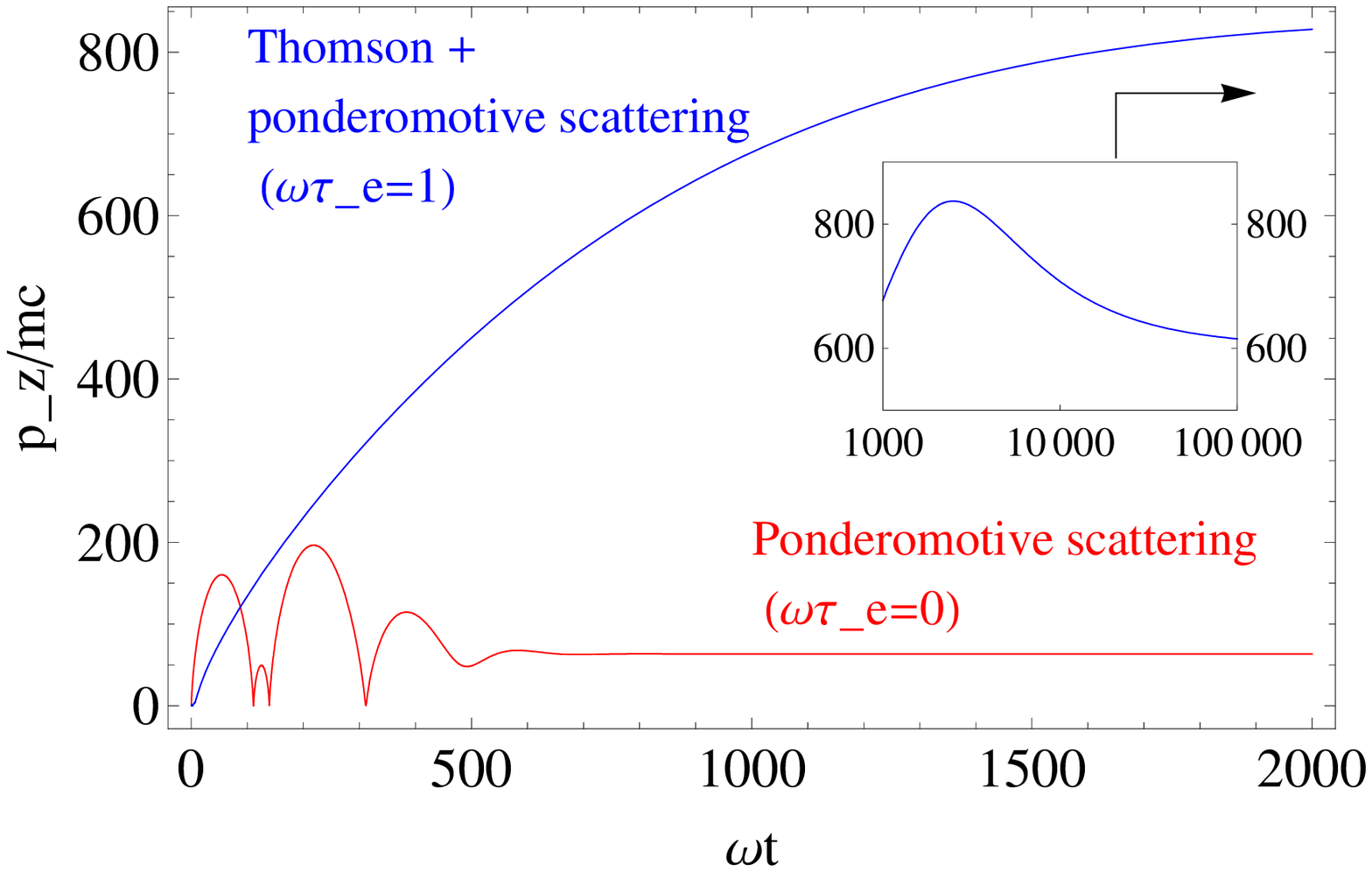}\\
\includegraphics[width=10cm]{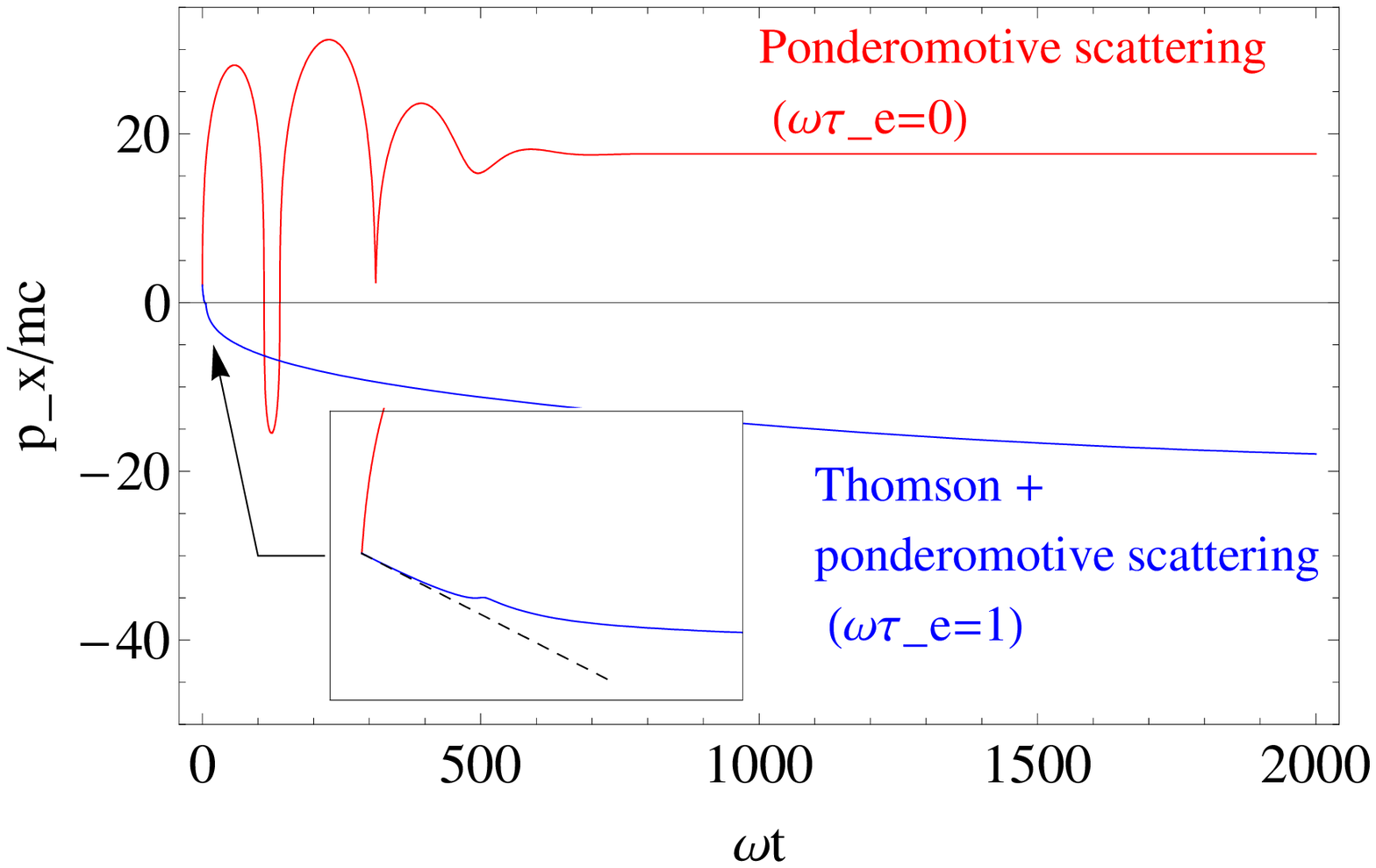}
\end{tabular}
\end{figure}

\clearpage
\begin{figure}[p!]
\caption{\label{fig4} \doublespacing
a) Trajectory of an electron bunch accelerated by a laser pulse with parameters $\lambda= 10$ $\mu$m, $I=1\cdot10^{19}$ W/cm$^2$, $\tau_L=2$ ps, and $w_0=10\lambda$; b) Longitudinal momentum of the bunch as a function of time; and
c) Transverse momentum of the bunch as a function of time, calculated for the case including radiation reaction (Thomson + ponderomotive scattering, $\omega\tau_e=1$) during the initial period $0\leq \omega t\leq 2\pi$ and for the case without radiation reaction (ponderomotive scattering, $\omega\tau_e=0$). The solid lines have been calculated using Eq. (\ref{8}) with $\bm{F}_L=q(\bm{E}+\bm{v}\times\bm{B})$, taking for $\bm{E}$ and $\bm{B}$ Eqs. (\ref{A1}-\ref{A4}). The inset of the middle panel shows the long-term behavior of the longitudinal momentum for the case including radiation reaction. The inset of the lower panel is a zoom-in on the first optical period; the slope of the dashed line is according to Eq. (\ref{14}). The initial position was $\bm{x}_0=(\lambda,0)$ and the initial velocity was $\bm{\beta}_0=0.9\bm{e}_x$.
}
\end{figure}

\end{document}